
\input phyzzx

\Pubnum{kcl-th-93-4}
\pubtype{}
\date{March 1993}

\titlepage

\title{A Covariant Canonical Description of Liouville Field Theory}

\author{G. Papadopoulos}
\address{Department of Mathematics \break King's College London\break
         London WC2R 2LS}
\andauthor{B. Spence}
\address{School of Physics\break University of Melbourne\break Parkville
          3052 Australia}

\abstract{We present a new parameterisation of the space of solutions of
Liouville field theory on a cylinder. In this parameterisation, the
solutions are well-defined and manifestly real
functions over all space-time and all of parameter space.  We show
that the resulting covariant phase space of the Liouville theory is
diffeomorphic to
the Hamiltonian one, and to the space of initial data of the theory.
The Poisson brackets are derived in our approach, and shown to
be those of the co-tangent bundle of the loop group of the real line.
Using Hamiltonian reduction, we show that our covariant phase space formulation
of Liouville theory may also be obtained from the covariant phase
space formulation of the Wess-Zumino-Witten
model that we have given previously. }

\endpage

\def\half{{1\over2}}

\def\pb#1#2{ \{#1,#2\} }
\def\pl{Phys. Lett.\ }


\def\exp{{\rm exp}}
\def\log{{\rm log}}
\def\dpl{\partial_+}
\def\dmi{\partial_-}
\def\p{\phi}

\def\intx{\int_01\!dx\,}
\def\sl2{sl(2,{\cal R})}
\def\eij{\epsilon_{ij}}
\def\l {\lambda}


\def\RN{{\cal R}}
\def\CN{{\cal C}}

\REF\liouville{J. Liouville, J. Math. Pure Appl. {\bf{18}} (1853) 71.}
\REF\polyakov{A.M. Polyakov, \pl {\bf B103} (1981) 207.}
\REF\curtright {T.L. Curtright and C.B. Thorn, Phys. Rev. Lett. {\bf{48}}
     (1982) 1309.}
\REF\braaten{E. Braaten, T.L. Curtright and C.B. Thorn, \pl {\bf{118B}} (1982)
    115; Ann. Phys. {\bf{147}} (1983) 365.}
\REF\jackiw {E. D'Hoker and R. Jackiw, Phys. Rev. {\bf D26} (1982) 3517.}
\REF\bonora {E. Aldrovandi, L. Bonora, V. Bonservizi and R. Paunov, {\it Free
     field representation of Toda field theories}, SISSA-ISAS 210/92/EP.}
\REF\gervaisa {J.-L Gervais and A. Neveu, Nucl. Phys. {\bf{B238}} (1984) 125;
     Nucl. Phys. {\bf B238} (1984) 396.}
\REF\gerv{J.-L. Gervais, LPTENS preprint 92/36, hepth/9212109.}
\REF\gawedski{K. Gawedski, talk presented at the
 Oji Seminar on Quantum Analysis,
             Kyoto 1992, hepth/9210100.}
\REF\weigt{G. Weigt, \pl {\bf B277} (1992) 79; DESY preprint 1992
      (talk presented at Johns Hopkins Workshop).}
\REF\babelon {O. Babelon, \pl {\bf 215B} (1988) 523.}
\REF\babelonb {O. Babelon, F. Toppan and L. Bonora, Commun. Maths. Phys. {\bf
     140} (1991) 93}
\REF\gervaisb {J.-L Gervais and A. Neveu, Nucl. Phys. {\bf B209} (1982) 125.}
\REF\leznov {A.N. Leznov and M.V. Saveliev, Lett. Math. Phys. {\bf 3} (1979)
      489.}
\REF\aldaya{V. Aldaya, J. Navarro-Salas and M. Navarro, CERN preprint TH.
            6393/92, to appear in Contemporary Mathematics.}
\REF\us{G. Papadopoulos and B. Spence, \pl {\bf 295B} (1992) 44.}
\REF\ustwo{G. Papadopoulos and B. Spence, {\it The canonical structure of
            the Wess-Zumino-Witten model}, to appear in the Proceedings of
             the NATO ASI Conference ``Low Dimensional Topology and
    Quantum Field Theory'', Newton Institute for Mathematical Sciences,
    Cambridge, September 1992.}
\REF\ellis {S.W. Hawking and G. F. R. Ellis, {\it The large scale structure of
   space-time}, CUP (1973).}
\REF\irish{P. Forg\'acs, A. Wipf, J. Balog, L. Feh\'er and
           L. O'Raifeartaigh, \pl {\bf 227B} (1989) 214.}

\sequentialequations


\chapter{Introduction}

The Liouville model [\liouville] is a venerable field theory, the study of
which has absorbed mathematicians and physicists for more than a century.
This theory has many applications in conformal field theory, 2d gravity and
particularily in the theory of non-critical strings [\polyakov].  Liouville
theory has a number of interesting classical and quantum properties.   For
example, there are B\"acklund transformations that relate the Liouville theory
to a free field theory [\curtright -\bonora] and  a quantum group structure
arises in this theory [\gervaisa -\babelonb].  The study of these
properties utilises an explicit parameterisation of the
solutions of the classical field equations of Liouville theory, and the
calculation of the Poisson brackets of the parameters of these
solutions.

The solutions $\phi$ of Liouville theory are functions from a
1+1-dimensional space-time into the real numbers $\RN$ and they are
parameterised by two independent parameters per space-time point.
In the literature there are two
distinct parameterisations of the solutions of
Liouville theory [\curtright-\jackiw, \gervaisb-\aldaya].  In the first,
the solutions are parameterised by two real (or
complex) independent functions $\{A,B\}$ of the light-cone
co-ordinates
$(x+, x-)\equiv (x+t, x-t)$ respectively.
In addition, for the Liouville theory on a cylinder an $SL(2,\RN)$ (or
$SL(2,\CN)$) monodromy $M$ can be
introduced that acts non-linearly on the parameters $\{A,B\}$.
In the second
approach, the space of solutions of Liouville theory is parameterised by
four real (or complex) functions $\{ui,vi\}$, $i=1,2$.
Both $u$ and $v$ transform under the
fundamental representation of $SL(2,\RN)$ (or $SL(2,\CN)$) and depend on the
light-cone co-ordinates $(x+, x-)$ respectively.
However, in this case two $SL(2,\RN)$ (or $SL(2,\CN)$)-invariant constraints
are introduced that depend on $\{u,v\}$ and their first derivatives. This
reduces
the number of independent parameters to two.  In this case the monodromy
$M$ acts linearly on $u$ and $v$.  The two distinct ways of parameterising
the space of solutions of Liouville theory can be related by constructing a
map from the $\{u,v\}$ to the $\{A,B\}$ parameters [\gervaisb].

In both of these parameterisations of the space of solutions of
Liouville theory, the relationship
between the parameters of the space of solutions
and the initial data of the theory is implicit (for a discussion of this point
see ref. [\gervaisb]). To our knowledge, no (explicit) diffeomorphism has been
constructed which relates the parameters of the solutions to the initial data
of
the Liouville theory.  Another problem that arises in the above
parameterisations is that for certain values of the parameters the solutions of
the Liouville theory become complex; this is the case even if the parameters
of the solutions are real functions. The reality condition on
the solutions imposes restrictions on their parameters.  These restrictions are
non-linear in the parameters and their first derivatives, and they affect the
quantum
mechanical treatment of the theory.  Furthermore, the solutions of the
Liouville theory expressed in the above parameterisations are not well-defined
on some regions of space-time for certain choices of parameters.  (For example,
in the $\{A,B\}$ parameterisation, the Liouville solution $e{-2\lambda\phi}
= (\partial_+A\partial_-B)/(1-\eta2AB)2$ ($\lambda$ and $\eta$ are coupling
constants) is not defined
at $t=0$ for $A=\eta2/B$.)  Similar problems also appear when the
solutions of the Liouville theory are expressed in terms of a free field and
they have been discussed by the authors of
refs. [\braaten]. The above problems are closely related to the difficulty of
specifying the {\it ranges} of the parameters $\{A,B\}$ and $\{u,v\}$ and their
first derivatives that parameterise the solutions of the Liouville theory.

In a recent paper [\us], see
also [\ustwo], we have given a new parameterisation of the space of
solutions of the Wess-Zumino-Witten (WZW) model with target space a compact and
connected group $G$, and we showed that the covariant and Hamiltonian phase
spaces of this theory are diffeomorphic.  The key point of this construction
was the introduction of a connection as part of the parameterisation of the
solutions of the WZW model. We showed how the parameters of our new
parameterisation of the WZW solutions are related
to the initial data of the theory by a simple transformation.
We finally used this new formulation to calculate the Poisson brackets
of the WZW theory in the covariant canonical approach.

In this paper, we will present a new parameterisation of the space of
solutions of the Liouville field theory on a cylinder.  As in the case of the
WZW model, this involves the introduction of a connection.
We will show that in this parameterisation the solutions of the Liouville
theory are well-defined functions from a cylinder (space-time) into the {\it
real} numbers.  We will then derive the Poisson brackets of the theory and
construct a symplectic diffeomorphism that relates the covariant phase space of
the Liouville theory to the Hamiltonian one, and to the space of initial data
of the theory.  Then we will show how these results can also be derived from
the Hamiltonian reduction of the covariant phase space of the WZW model which
we have given in ref. [\us].


\chapter{Covariant Canonical Description of Liouville Theory}

 The phase space of a classical system can be defined in two
different ways.  The first gives the Hamiltonian phase space $P_H$, which is
defined as the space of positions and
momenta of a system, equipped with the standard symplectic form.  For the
second
definition, we begin with the space of fields of a system.  Then we introduce
the symplectic current $S{\mu}=\delta \phiI \delta \big(\partial
L/\partial(\partial_{\mu}\phiI)\big)$, where $\phi$ is the field
and $L$ is the Lagrangian of the theory.  This current is conserved
($\partial_{\mu} S{\mu}=0$) when the
Lagrangian equations of motion of the model are satisfied.
The Lagrangian symplectic form $\Omega$ is
defined as the integral over a Cauchy surface of the time component of the
symplectic current.  This two-form is closed and independent of the choice of
the Cauchy surface that we have used to define it. This symplectic form can
be parameterised in different ways.  One way is in terms of the initial data of
the theory.  In theories for
which we know an explicit parameterisation of the space
of solutions, the Lagrangian symplectic form can also
be parameterised directly in terms
of the parameters of the solutions.  The covariant phase space $P_C$ of a
theory is then defined as
the space of solutions of the Lagrangian equations of motion of the
system, equipped with the Lagrangian symplectic form written in terms of the
parameters of the solutions.

The Lagrangian of the Liouville theory is
$$
L=-2(\partial_+\p \dmi\p +{\eta2\over \l2} e{-2 \l \p}),
\eqn\aone
$$
where $\p$ is a map from a cylinder $S1\times {\RN}$ to ${\RN}$, and $\l$ and
$\eta$ are real coupling constants. The pairs
$(x,t): 0\leq x<1, -\infty<t<\infty$ are the
co-ordinates of $S1\times {\RN}$ and  $x\pm = x \pm t, \partial_\pm =
\half(\partial_x \pm\partial_t)$. The equations
of motion are
$$
     \dpl\dmi\p + {\eta2\over \l}e{-2 \l \p}=0.        \eqn\one
$$
 The Lagrangian symplectic form of the model is
$$
    \Omega = \intx\delta\p\,\partial_t\delta\p,      \eqn\two
$$
evaluated at $t=0$.

To construct a new parameterisation of the space of solutions of the Liouville
theory, we set
$$
 e{\lambda\p (x,t)} =
\epsilon_{ij}ui(x+){Wj}_l(A;x+,x-)vl(x-),
 \eqn\three
$$
where indices $i,j,l=1,2$, $\epsilon$ is the
$2\times 2$ anti-symmetric matrix with $\epsilon_{12} = 1$,
$u$ and $v$ are periodic functions from the real line $\RN$ into ${\RN}2$,
and
$W(A;x+,x-)=\{{Wj}_l(A;x+,x-)\}$ is the holonomy of a connection $A$
given by
$$
      W(A;x+,x-) =  P\,\exp\int_{x-}{x+} A(s) ds.    \eqn\four
$$
In the above, $A$ is a periodic one-form on the real line, with values into
$(LieSL(2,\RN))*$ the dual of the Lie algebra of $SL(2,{\RN})$. Note that we
take $u$ to be a function of
$x+$, and
$v$ a function of $x-$. Finally, the functions $u,v$ are
required to satisfy the constraints
$$
     \eij ui \nabla_+ uj = -\nu, \quad  \eij  vi \nabla_-vj =
{\eta2\over \nu},
                                                      \eqn\five
$$
where the covariant derivatives are given by
$$
     \nabla_+ ui = \dpl ui - {Ai}_j uj, \quad
      \nabla_- vi = \dmi vi - {Ai}_j vj,             \eqn\six
$$
and $\nu$ is a real constant ($\nu\not=0$).
We note that the expression for $\p$ in equation \three\ above, and
the constraints \five, are invariant under the gauge transformations
$$
     \eqalign { & u(x)\rightarrow h{-1}(x)u(x),
      \quad v(x)\rightarrow h{-1}(x)v(x),
         \cr &A(x)\rightarrow -\partial_xh(x)\,h{-1}(x) +
           h(x)A(x)h{-1}(x),     \cr}       \eqn\seven
$$
where $h$ is an element of the loop group $LSL(2,{\RN})$ of $SL(2,{\RN})$.

It can be shown from the periodicity of $u,v$ and $A$ that $\p$ is
periodic in $x$.  In addition, if  $u,v$ and $A$ satisfy the conditions of
eqn. \five, then
$\p$ of eqn. \three\ solves the equations of motion of Liouville theory, eqn.
\one.  However note that $\phi$ in equation \one\ is not necessarily a real
function.  As we will explain later, the reality condition on the solutions
$\phi$ imposes additional restrictions on the parameters $u,v,A$.

This method of parameterising the solutions of the Liouville theory
is similar to (and was inspired by) the one
constructed in ref. [\us] for the WZW model. Note, however, that in this case
there are the conditions of eqn. \five\  together with the requirement
that the solutions $\phi$ are real functions. These are additional restrictions
on the parameters of the theory.
Because of this, we will not treat the above
construction as fundamental as we did in the case of the WZW model.
Rather, we will consider
it as a device to give a parameterisation to the space of
solutions of Liouville theory.

In order to express the solutions of the field equations in terms of
independent parameters and calculate the symplectic form of the covariant phase
space of the Liouville theory, we first have to gauge fix the symmetry
generated by the group action of
eqn. \seven\ and then solve the conditions of eqn. \five.

Due to eqn. \five, the parameters $u,v$ are not allowed to be zero.
To obtain a non-degenerate symplectic
form on $P_C$ we must fix the gauge symmetry. We will choose the following
gauge: writing $v = \pmatrix{v1\cr v2\cr}$, at least one of $v1, v2$
must be non-zero. If $v2\not=0$, then we choose $u=u_{{}_0}=\pmatrix{1\cr
0\cr}$,
and $v=v_{{}_0}= k \pmatrix{0\cr 1\cr}$, where $k$ is a map from $S1$ into the
real line minus the origin $\RN-\{0\}$ (\ie\ $k\not=0$).
In fact, we will show later (by comparison with the initial
data) that $k$ should take values in the {\it positive} real line
$\RN+-\{0\}$.
If $v2=0$ (and so $v1\not=0$), then we choose $u=\pmatrix{0\cr 1\cr}$,
and $v= k \pmatrix{1\cr 0\cr}$.

We now consider the first gauge fixing, \ie\ assume that $v2\not=0$
(analogous results obtain for the other case).
Then we pick the following basis in $(Lie SL(2,{\RN}))*$:
$$
t{{}0}=\{{(t{{}0})i}_j\} = \pmatrix {1&0\cr 0&-1\cr},\quad
t+=\{{(t+)i}_j\} = \pmatrix {0&1\cr 0&0\cr},\quad
t-=\{{(t{-})i}_j\} = \pmatrix {0&0\cr 1&0\cr}.  \eqn\eightb
$$
Substituting the gauge conditions
into the constraints \five\ and solving them in terms of the components of
$A=A_{{}_0}t{{}0}+A_+ t++A_-t-$, we get
$$
   A_- = \nu,\quad A_+ ={\eta2\over \nu} {1\over k2}.   \eqn\nine
$$
Thus, the independent variables after gauge-fixing are $k$ and the component
$A_{{}_0}$ of the connection $A$. Substituting the gauge-fixing conditions and
the
solutions \nine\ of the conditions \five\ into the parameterisation
 \three, we get a
parameterisation of the space of solutions of the Liouville theory in terms of
$k$ and $A_{{}_0}$.  This is
$$\eqalign{ e{\lambda\p (x,t)} &=
\epsilon_{ij}u_{{}_0}{}i{Wj}_l(a;x+,x-)v_{{}_0}{}l(x-)  \cr
        &= W2{}_2(a;x+,x-) k(x-),  \cr}
\eqn\aanine
$$
where $a=A_{{}_0}t{{}0}+{\eta2\over \nu} {1\over k2}t+ +\nu t-$.
In addition, if we require that $k$ takes values into the positive real line,
then the solutions
$\phi$ (eqn.\aanine) of the Liouville theory parameterised in terms of
$A_{{}_0}$ and $k$ are functions from a cylinder into the real line $\RN$, \ie\
they are well-defined and real  at every point of the space time for any choice
of the parameters $A_{{}_0}$ and $k$ (we will justify this at the end of this
section).  The space of parameters of
the solutions of the Liouville theory is the co-tangent bundle of
the loop group $L(\RN+-\{0\})$ and it is diffeomorphic to the co-tangent
bundle of $L\RN$.

The Lagrangian symplectic form \two\ parameterised in
terms of $k,A_{{}_0}$ is then found to be
$$
 \Omega =-{1\over \l2}  \intx \bigg( 2\,\delta \log (k)\delta A_{{}_0}+\delta
\log (k)
\,\partial_x\delta \log (k)\bigg).    \eqn\ten
$$
This symplectic form is not degenerate and can be easily inverted.  The Poisson
brackets of the Liouville theory are thus found to be
$$
\eqalign{ \pb{k(x)}{k(y)} & = 0, \cr
          \pb{k(x)}{A_{{}_0}(y)} & = -{\l2\over 2} k(x)\delta(x,y),\cr
          \pb{A_{{}_0}(x)}{A_{{}_0}(y)} & = {\l2\over 2}
{\partial_x}\delta(x,y),\cr}
     \eqn\eleven
$$
where $\delta(x,y)$ is the delta function on a circle. Note that in the
variables $q,p$ defined by
$q = \log k$ and $p = -\partial_x\log k - 2 A_{{}_0}$,
the symplectic form is simply $\Omega = {1\over \l2}\intx \delta q\delta p$.
These Poisson brackets
are those of the co-tangent bundle of the loop group $L\RN$ of $\RN$.
In the next section, we will see how this may be understood using the
Hamiltonian reduction approach.

Next we will compare the initial data of the Liouville theory with the
parameters $k,A_{{}_0}$ of the solutions of the theory. The Lagrangian
symplectic
form \two\ parameterised in terms of the initial data $f(x)=\lambda\phi(x,0)$
and
$w(x) = \lambda\partial_t\phi(x,0)$, is
$\Omega = {1\over \l2}\intx\delta f\delta w$
and the associated Poisson brackets are $\pb{f(x)}{w(y)} = {\l2} \delta(x,y)$.
 The space of initial data with this symplectic structure is
isomorphic to the Hamiltonian phase space $P_H$ of the Liouville theory and the
initial data can be identified with the positions and momenta of the theory.
Now we can easily express the initial data of the Liouville theory in terms of
the parameters $k,A_{{}_0}$. Indeed,
$$(f,w)=(\log k,-\partial_x\log k- 2 A_{{}_0}).\eqn\atwelve$$
Since $f$ is a {\it real} function, the parameter $k$ is required to take
values on
the {\it positive} real line. Under this requirement, map given in eqn.
\atwelve\ is a diffeomorphism, in fact a
{\it symplectic} diffeomorphism, as the brackets \eleven\ of $P_C$ become
precisely the canonical brackets of $P_H$ under this map.
Thus we conclude that
the covariant $P_C$ and Hamiltonian $P_H$ phase spaces of the Liouville
theory are isomorphic.

The solutions of the Liouville theory parameterised in terms of $A_{{}_0}$ and
$k$ ($k>0$) are {\it real}.  For  solutions analytic in the time co-ordinate
$t$, this
can be  shown explicitly by expressing these solutions as power series in $t$.
For solutions which are not analytic in $t$,
we note that the Liouville theory has a
well-posed initial value problem [\ellis], which amongst other things imples
that there is a unique solution of the Liouville equation for every set of
initial data $(f,w)$.  Now we observe that if $\phi$ is a solution of the
Liouville equation and $\phi$ is complex, then its complex conjugate $\phi*$
is
a solution as well.  However if the initial data $(f,w)$ ($k>0$) of $\phi$ are
real, it is easy to prove that both $\phi$ and $\phi*$ have the same initial
data.  Hence they must be equal  ($\phi=\phi*$), and $\phi$ is real.

Finally, the solutions of the Liouville theory parameterised in terms of
$A_{{}_0}$ and $k$  can be re-written in terms of variables $n$ and $m$ such
that $n$ depends on $x+$ and $m$ depends on $x-$ correspondingly.  To do
this, we choose a point $x_{{}_0}$ on the real line and write the solution
as
$e{\lambda \p(x,t)}=\eij ni(x+,x_{{}_0}) mj(x-,x_{{}_0})$
where
$n(x,x_{{}_0})=W(a;x_{{}_0},x) u_{{}_0}$  and $m(x,x_{{}_0})=W(a;x_{{}_0},x)
v_{{}_0}$.
The solution $\p$ does not depend on the choice of $x_{{}_0}$.


\chapter {Hamiltonian Reduction}

An alternative approach to Liouville theory considers it as a Hamiltonian
reduction of the Wess-Zumino-Witten (WZW) model with target space the group
$SL(2,{\RN})$ [\irish]. We will
now show that this reduction applies to the covariant canonical approach,
in that the Hamiltonian reduction of the covariant canonical description
of the WZW model, as given in ref. [\us], yields the covariant canonical
description of the Liouville theory presented above.

Firstly we recall some results from ref. [\us]. The parameterisation of
the classical solutions of the Lagrangian equations of motion of the WZW model
on a
cylinder with target space a
compact and connected Lie group $G$ is
$$
\eqalign{   g(x,t) &= U(x+) {\cal W}(A;x+,x-) V(x-), \crr
                 {\cal W}(A;x+,x-) &=
                     P\exp\!\int_{x-}{x+}\!A(s)ds,\cr } \eqn\thirteen
$$
where $U$ and $V$ are periodic maps from the real line ${\RN}$ to the
group $G$, and the field $A$ in the path-ordered exponential is a
 $({\rm Lie}G)*$-valued periodic one-form on
the real line. The expression for $g(x,t)$ in eqn. \thirteen\ is then
periodic in $x$ and solves the field equations
$$
   \partial_-(\partial_+  g\ g{-1})=0, \eqn\fourteen
$$
where $g$, a map from the cylinder into $G$, is the WZW field.
The parameterisation \thirteen\ has the symmetry
$$
     \eqalign { & U(x)\rightarrow U(x)h(x),
          \quad V(x)\rightarrow h{-1}(x)V(x),
         \cr &A(x)\rightarrow -h{-1}(x)\partial_xh(x) +
           h{-1}(x)A(x)h(x),     \cr}       \eqn\fifteen
$$
where $h$ is an element of the loop group of $G$. These results can be extended
to include the WZW model whose target space is the group
$SL(2,{\RN})$.

Next we will show that a Hamiltonian
reduction of this parameterisation of the space of solutions of the WZW
model gives the parameterisation of the space of solutions of the
Liouville field theory which we gave in the previous section.

The currents of the WZW model are
given by
$$
      J_+(x+) =-{\kappa\over 4\pi} \partial_+  g\ g{-1},\quad
        J_-(x-) = {\kappa\over 4\pi} g{-1}\partial_- g.  \eqn\fifteena
$$
where $\kappa$ is a coupling constant.
Hamiltonian reduction of the WZW model with target space the group
$SL(2,{\RN})$
to the Liouville theory occurs when one imposes the
constraints
$$
 J_+|_{t-} = -\nu {\kappa\over 4\pi}, \qquad  J_-|_{t+} = -{\kappa \eta2
\over 4\pi \nu},
 \eqn\sixteen
$$
on the components of the currents of the WZW model.
Rewriting the constraints \fourteen\ in terms of the variables $U, V, A$ of the
parameterisation
\fourteen, they become
$$
   \eqalign{
  \big(\partial_+U U{-1}+ UA(x+)U{-1})\big)|_{t-} &= \nu,\cr
    \big(V{-1}\partial_-V -V{-1}A(x-)V\big)|_{t+} &= -{\eta2\over
\nu}.\cr}
    \eqn\seventeen
$$
Now we introduce $ui, vi$ satisfying the constraints \five. Then it is
straightforward to show that the $SL(2,{\RN})$ matrices
$$
   U = \pmatrix{{u1\over M} &{u2\over M} \cr - u2 & u1 \cr}, \quad
   V = \pmatrix{{v2\over N} & v1 \cr - {v1\over N} & v2 \cr},
\eqn\eighteen
$$
 satisfy
the constraints \seventeen\ where $M = {(u1)}2 + {(u2)}2, N =
{(v1)}2 + {(v2)}2$. Notice that $M$ and $N$ cannot be zero for $ui,
vi$ that satisfy the condition \five. Following the arguments of ref.
[\irish],
it then follows that the Liouville field $\phi$ is given by
$$
      e{\l \phi(x,t)}\equiv (UWV)2{}_2 =
{\Big(U(x+)W(A;x+,x-)V(x-)\Big)}2{}_2 =
     \epsilon_{ij}ui(x+)Wj{}_l(A;x+,x-) vl(x-),
    \eqn\nineteen
$$
where $(UWV)2{}_2 $ is the $(22)$ component of the $SL(2,{\RN})$ matrix
$UWV$.
Thus, the Hamiltonian reduction defined by equation \sixteen\ reduces
the WZW parameterisation \thirteen\ to the Liouville parameterisation
\three.  However, this Hamiltonian reduction does not necessarily lead to
solutions $\phi$ of the Liouville theory that are real functions.  The reality
condition on these solutions arises as an additional restriction on their
parameters and can be handled as in the previous section.

To show that the reduction is indeed Hamiltonian, \ie\ that the
WZW Poisson brackets reduce to the Liouville Poisson brackets, we recall
that in ref. [\us] we showed that the
covariant canonical phase space of the WZW model was the co-tangent bundle of
the loop group of $G$, agreeing with the Hamiltonian phase space. The
brackets are the canonical brackets on this phase space. This result
also applies to the case $G = SL(2,{\RN})$, and thus we have the result
that the Poisson
brackets of the  WZW model with target space group $SL(2,{\RN})$ in the gauge
$U=1$ are
$$
\eqalign{ \pb{V(x)}{V(y)} & = 0, \cr
          \pb{V(x)}{A_a(y)} & =\beta V(x)t_a\delta(x,y),\cr
          \pb{A_a(x)}{A_b(y)} & = \beta \big(\delta_{ab}
         {\partial\over\partial x}\delta(x,y) +
             {f_{ab}}cA_c(x)\delta(x,y)\big),\cr}
     \eqn\thirteen
$$
where $V(x)$ is an element of $LSL(2,{\RN})$ and $\beta=-{4\pi\over \kappa}$.
The $t_a$ are the generators of $Lie(SL(2,{\RN}))*$, satisfying $[t_a,t_b] =
{f_{ab}}ct_c$, with ${f_{ab}}c$ the structure constants of $SL(2,{\RN})$.
If we express $V$ in terms of $v$ as in eqn. \eighteen, gauge fix $u$ and $v$
as in the previous section and  use the algebra  $[t{{}0},t\pm] =\pm 2
t\pm,
[t+,t-] = t{{}0}$ of the basis $t{{}0},t+,t-$, the Poisson
brackets of eqn. \thirteen\ reduce to the Poisson brackets of eqn. \eleven\
provided that $\beta=\half\l2$.  Thus we see that the covariant canonical
formulation
of the Liouville theory presented in the previous section
can be derived by a Hamiltonian reduction of
the covariant canonical formulation of the WZW model which we gave in
ref. [\us].

\chapter {Conclusions}
The parameterisation described in section two can also be used to study the
Liouville theory on a flat two-dimensional Minkowski space.  The analysis for
parameterising the solutions can be done in the same way as for the Liouville
theory on the cylinder.  However in this case, $u,v,A$ are not necesarily
periodic.  As a result the parameters $A_{{}_0}$ and $k$ are smooth functions
from the real line $\RN$ into the real line $\RN$.  The calculation of the
covariant symplectic form and the associated Poisson brackets can formally
proceed as in the case of the Liouville theory on a cylinder.  However, it
should be noted that the covariant symplectic form is not well defined for all
smooth $A_{{}_0},k, \delta A_{{}_0},\delta k$.

To summarise, we have presented a parameterisation of the space of
solutions of the Liouville theory on a cylinder with the following properties:
The solutions of the Liouville theory are functions from a cylinder
(space-time) into the real numbers.  We showed that the space of parameters of
the solutions is diffeomorphic to the space of initial data of the theory and
that this diffeomorphism induces a symplectic diffeomorphism from the covariant
phase space onto the Hamiltonian phase space of the Liouville theory. The
Poisson brackets of the Liouville theory were also derived.  Finally, we
showed that our covariant phase space description of the Liouville theory
can be also derived, via Hamiltonian reduction, from the covariant phase space
description of the WZW model given by us in ref. [\us].

\noindent{\bf Acknowledgements:} G.P. was supported by the Commission of
European Communities, and B.S.
by a Queen Elizabeth II Fellowship from the Australian Government.

\refout

\bye